\documentclass[twocolumn]{aastex631}

\usepackage{CJK}
\CJKspace
\usepackage{amsmath}

\newcommand{\msun}{$M_\mathrm{\odot}$}
\newcommand{\msunyr}{$M_\mathrm{\odot~yr^{-1}}$}
\newcommand{\Ha}{$\mathrm{H\alpha}$}
\newcommand{\Hb}{$\mathrm{H\beta}$}
\newcommand{\MgIIlambda}{\ion{Mg}{2} $\lambda\lambda$2796, 2802}
\newcommand{\MgII}{\ion{Mg}{2}}
\newcommand{\kms}{$\mathrm{km~s^{-1}}$}
\newcommand{\pwrunit}{$\mathrm{erg~s^{-1}}$}
\newcommand{\densityunit}{$\mathrm{g~cm^{-3}}$}
\newcommand{\masslossunit}{$M\mathrm{_{\odot}~yr^{-1}}$}

\usepackage{soul}

\defcitealias{astropy_collaboration_astropy_2013}{Astropy Collaboration 2013}
\defcitealias{astropy_collaboration_astropy_2018}{2018}
\defcitealias{astropy_collaboration_astropy_2022}{2022}

\newcommand{\LCO}{\affiliation{Las Cumbres Observatory, 6740 Cortona Drive, Suite 102, Goleta, CA 93117-5575, USA}}
\newcommand{\UCSB}{\affiliation{Department of Physics, University of California, Santa Barbara, CA 93106-9530, USA}}
\newcommand{\UCSD}{\affiliation{Department of Astronomy \& Astrophysics, University of California, San Diego, 9500 Gilman Drive, MC 0424, La Jolla, CA 92093-0424, USA}}

\newcommand{\UCD}{\affiliation{Department of Physics and Astronomy, University of California, Davis, 1 Shields Avenue, Davis, CA 95616-5270, USA}}

\newcommand{\GSFC}{\affiliation{Astrophysics Science Division, NASA Goddard Space Flight Center, Mail Code 661, Greenbelt, MD 20771, USA}}

\newcommand{\UCB}{\affiliation{Department of Astronomy, University of California, Berkeley, CA 94720-3411, USA}}

\newcommand{\STScI}{\affiliation{Space Telescope Science Institute, 3700 San Martin Drive, Baltimore, MD 21218-2410, USA}}

\newcommand{\IPAC}{\affiliation{IPAC, Mail Code 100-22, Caltech, 1200 E.\ California Blvd., Pasadena, CA 91125}}

\newcommand{\CfA}{\affiliation{Center for Astrophysics \textbar{} Harvard \& Smithsonian, 60 Garden Street, Cambridge, MA 02138-1516, USA}}
\newcommand{\UA}{\affiliation{Steward Observatory, University of Arizona, 933 North Cherry Avenue, Tucson, AZ 85721-0065, USA}}

\newcommand{\GeminiNorth}{\affiliation{Gemini Observatory, 670 North A`ohoku Place, Hilo, HI 96720-2700, USA}}
\newcommand{\Keck}{\affiliation{W.~M.~Keck Observatory, 65-1120 M\=amalahoa Highway, Kamuela, HI 96743-8431, USA}}

\newcommand{\Catalyst}{\altaffiliation{LSST-DA Catalyst Fellow}}

\newcommand{\Rutgers}{\affiliation{Department of Physics and Astronomy, Rutgers, the State University of New Jersey,\\136 Frelinghuysen Road, Piscataway, NJ 08854-8019, USA}}

\newcommand{\IAP}{\affiliation{Institut d'Astrophysique de Paris, CNRS-Sorbonne Universit\'e, 98 bis boulevard Arago, 75014 Paris, France}}

\newcommand{\IAIFI}{\affiliation{The NSF AI Institute for Artificial Intelligence and Fundamental Interactions, USA}}

\begin{document}

\title{Circumstellar Interaction in the Ultraviolet Spectra of SN~2023ixf 14--66 Days After Explosion}
\correspondingauthor{K. Azalee Bostroem}
\email{bostroem@arizona.edu}
\author[0000-0002-4924-444X]{K.\ Azalee Bostroem}
\Catalyst\UA
\author[0000-0003-4102-380X]{David J.\ Sand}
\UA
\author[0000-0003-0599-8407]{Luc Dessart}
\IAP
\author[0000-0001-5510-2424]{Nathan Smith}
\UA
\author[0000-0001-8738-6011]{Saurabh W.\ Jha}
\Rutgers
\author[0000-0001-8818-0795]{Stefano Valenti}
\UCD
\author[0000-0003-0123-0062]{Jennifer E.\ Andrews}
\GeminiNorth
\author[0000-0002-7937-6371]{Yize Dong \begin{CJK*}{UTF8}{gbsn}(董一泽)\end{CJK*}}
\UCD
\author[0000-0003-3460-0103]{Alexei V. Filippenko}
\UCB
\author[0000-0001-6395-6702]{Sebastian Gomez}
\STScI
\author[0000-0002-1125-9187]{Daichi Hiramatsu}
\CfA\IAIFI
\author[0000-0003-2744-4755]{Emily T. Hoang}
\UCD
\author[0000-0002-0832-2974]{Griffin Hosseinzadeh}
\UCSD
\author[0000-0003-4253-656X]{D.\ Andrew Howell}
\LCO\UCSB
\author[0000-0001-5754-4007]{Jacob E.\ Jencson}
\IPAC
\author[0000-0001-9589-3793]{Michael Lundquist}
\Keck
\author[0000-0001-5807-7893]{Curtis McCully}
\LCO\UCSB
\author[0009-0008-9693-4348]{Darshana Mehta}
\UCD
\author[0000-0002-7015-3446]{Nicolas E.\ Meza Retamal}
\UCD
\author[0000-0002-0744-0047]{Jeniveve Pearson}
\UA
\author[0000-0002-7352-7845]{Aravind P.\ Ravi}
\UCD
\author[0000-0002-4022-1874]{Manisha Shrestha}
\UA
\author[0000-0003-2732-4956]{Samuel Wyatt}
\GSFC

\begin{abstract}
SN~2023ixf was discovered in M101 within a day of explosion and rapidly classified as a Type II supernova with flash features. 
Here we present ultraviolet (UV) spectra obtained with the {\it Hubble Space Telescope} 14, 19, 24, and 66 days after explosion. 
Interaction between the supernova ejecta and circumstellar material (CSM) is seen in the UV throughout our observations in the flux of the first three epochs and asymmetric \MgII\ emission on day 66. 
We compare our observations to CMFGEN supernova models that include CSM interaction ($\dot{M}<10^{-3}$ \masslossunit) and find that the power from CSM interaction is decreasing with time, from $L_{\rm sh}\approx5\times10^{42}$ \pwrunit\ to $L_{\rm sh}\approx1\times10^{40}$ \pwrunit\ between days 14 and 66. 
We examine the contribution of individual atomic species to the spectra on days 14 and 19, showing that the majority of the features are dominated by iron, nickel, magnesium, and chromium absorption in the ejecta.
The UV spectral energy distribution of SN~2023ixf sits between that of supernovae which show no definitive signs of CSM interaction and those with persistent signatures assuming the same progenitor radius and metallicity. 
Finally, we show that the evolution and asymmetric shape of the \MgIIlambda\ emission are not unique to SN~2023ixf. 
These observations add to the early measurements of dense, confined CSM interaction, tracing the mass-loss history of SN~2023ixf to $\sim33$ yr prior to the explosion and the density profile to a radius of $\sim5.7\times10^{15}$ cm. 
They show the relatively short evolution from a quiescent red supergiant wind to high mass loss. 
\end{abstract}


\section{Introduction}
Hydrogen-rich supernovae (Type II) are thought to come from red supergiants (RSGs) with zero-age main sequence masses between $\sim8$ and $\sim25$ \msun. 
While it is accepted that all RSGs lose mass, there are no quantitative mass-loss rates derived from stellar evolution theory. 
Adding to this, there is no consensus about the empirical mass-loss rates for RSGs; \citet{2023massey} infer mass-loss rates higher than the canonical rates of \citet{1988deJager} from the RSG luminosity function, while other studies \citep[e.g.][]{2020Beasor}, measure lower mass-loss rates from the infrared (IR) excess of RSGs. 
Different mass-loss rates affect the final star that explodes, impacting the final hydrogen envelope mass as well as the density structure of the circumstellar material (CSM) into which the supernova explodes.

Signatures of mass loss can also be observed in supernovae, allowing us to constrain late-stage RSG mass loss using post-explosion observations.
One definitive sign of CSM is the presence of narrow emission lines. 
As the number of young Type II supernovae discovered has increased with the prevalence of high-cadence wide-field surveys, it has become clear that a significant fraction of them have confined CSM \citep{2016Khazov,  2023Bruch, 2024Jacobson-Galan}. 
Early-time spectra of these supernovae display flash features: narrow emission lines with Lorentzian wings formed by the ionization of surrounding CSM by shock breakout and the interaction of the ejecta with CSM.
These narrow features disappear a few days to weeks after explosion and the evolution then proceeds as a typical Type IIP/L supernovae.

While these lines disappear when the CSM is swept up, the influence of this CSM on the remaining photometric and spectroscopic evolution of the supernova is uncertain.
Observations and recent theoretical modeling of the effects of the overall dynamics and CSM properties on the light curve and spectral evolution suggest that, even in the absence of narrow emission lines, the CSM-ejecta interaction will convert some of the kinetic energy of the supernova into thermal energy, boosting the supernova luminosity \citep{2015Smith, 2017Dessart, 2022Dessart}. 
Furthermore, this interaction will produce a bluer spectral energy distribution (SED) after a few days. 
The discrepancy between the $U-V$ color curves of a supernova with and without CSM grows with time as the noninteracting supernova ejecta expand and cool \citep{2022Dessart}.
Spectroscopically, high-velocity absorption features, from the cool dense shell that forms at the interface between ejecta and CSM, are predicted to be visible in \Ha, \Hb\ \citep{2007Chugai}, \ion{Na}{1\,D} $\lambda\lambda$5890, 5896, and the \ion{Ca}{2} near-infrared (NIR) triplet $\lambda\lambda\lambda$8498, 8542, 8662, and produce strong emission lines in the ultraviolet \citep{2022Dessart}. 
In particular, as the ejecta cool, \MgIIlambda{} emerges in emission with increasing strength as the level of CSM interaction increases.
Thus, the characterization of the long-term evolution of supernovae weeks to years after explosion probes RSG mass loss centuries to millennia prior to explosion.

Observational evidence supporting these predictions is unclear. Sample analyses found contradictory results about whether supernovae with flash features are systematically bluer and more luminous than noninteracitng supernovae \citep{2016Khazov, 2023Bruch}. 
Additional contradictions are found in individual objects (e.g. SN~2018zd, \citealt{2021Hiramatsu}; SN~2020pni, \citealt{2022Terreran}; various, \citealt{2024Jacobson-Galan}; PTF11iqb, \citealt{2015Smith}; SN~2013fs, \citealt{2018Bullivant}).
Spectroscopically, while an absorption feature is often observed blueward of \Ha{} \citep{2017Gutierrez} and has been interpreted as high-velocity hydrogen due to CSM interaction, the relationship between the sample of supernovae with this feature and those with flash features is unclear.
UV observations of Type II supernovae are especially sparse, with early-time data limited by the availability of sensitive rapid-response UV telescopes, and with late-time observations limited by the fading of the UV when CSM interaction is not present \citep{1980Panagia,1982Pettini,1987Cassatella2,2000Baron,2005Fransson,2022Vasylyev, 2023Vasylyev}.

The study of Type II supernovae with dense, confined CSM was recently enriched by the explosion of SN~2023ixf in the nearby galaxy M101. 
The proximity, early discovery and classification \citep{itagaki_discovery_2023, perley_classification_2023}, and identification of narrow emission lines have led to this being the best-observed supernova with flash features to date. 
Of particular note is the periodic pre-explosion light curve of the RSG progenitor in the optical and IR \citep[][]{2023Jencson, 2023Kilpatrick, 2023Niu, 2023Qin, 2023Soraisam,2024Xiang}, the incredibly deep multiband pre-discovery data which characterize the shock breakout and light-curve rise starting at an absolute magnitude of $V \approx -11$ together with high-cadence photometry through the end of the plateau phase \citep{2023Jacobson-Galan, 2023Hiramatsu, 2023Hosseinzadeh,  2023Sgro, 2023Teja, 2023Yamanaka, 2023Zimmerman,  2024Bersten, 2024Li, 2024Martinez}, intranight optical spectroscopy for more than a week after discovery \citep{2023Bostroem2, 2023Jacobson-Galan,2023Teja, 2023Hiramatsu, 2023Yamanaka, 2023Zhang}, early-time optical spectropolarimetry \citep{2023Vasylyev2, 2024Singh}, high-resolution spectroscopy of the first 17 days of evolution \citep{2023Smith, 2023Hiramatsu, 2023Zimmerman},
X-ray and radio detections \citep{2023Berger,2023Grefenstette,2023Matthews, 2024Chandra}, and the first early-time UV spectra of a supernova  with flash features \citep{2023Zimmerman}.
While a coherent picture of these observations is still being explored, they point toward a confined, dense CSM with a radius $R\lesssim10^{15}$ cm and a mass-loss rate $\dot M \approx 10^{-2}$--$10^{-4}$  \msunyr. 

In this paper, we describe UV spectroscopy of SN~2023ixf obtained with the {\it Hubble Space Telescope (HST)} at 14, 19, 24, and 66 days past explosion, all after the flash features had faded. 
Throughout the paper we use UTC 2023-05-18 18:00:00 \citep[MJD 60082.75 $\pm$ 0.10;][]{2023Hosseinzadeh} as the explosion epoch, $E(B-V) = 0.0077 \pm 0.0002$ mag \citep{2011Schlafly} as the Milky Way (MW) extinction, and $E(B-V) = 0.031$ mag \citep{2023Smith} as the host-galaxy extinction. 
In \autoref{sec:data} we describe the data reduction and in \autoref{sec:SpecEvolve} the evolution of the spectra, focusing on the UV.
SN~2023ixf is compared to models with CSM interaction in \autoref{sec:CompModel} and to other supernovae with UV observations in \autoref{sec:CompObs}.
We close with a discussion of CSM interaction through day 66 in \autoref{sec:discussion} and summarize our results in \autoref{sec:summary}.

\section{Observations and Data Reduction}\label{sec:data}
{\it HST} observations were obtained on days 14, 19, 24, and 66 with the CCD detector of the Space Telescope Imaging Spectrograph (STIS) using the 52\arcsec $\times$ 0.\arcsec2 slit at the E1 position to mitigate flux loss to charge transfer efficiency loss and the G230LB, G430L, and G750L gratings \citep[DD GO-17313; ][]{2023Bostroem3}.
Unfortunately, our planned visit on day 50 failed to acquire guide stars. 
However, the observation was successfully repeated on day 66.
We obtained at least four exposures with each grating to automatically remove cosmic rays. 
A list of observations is given in \autoref{tab:HSTspec}.

Bias, flat-fielding, and cosmic-ray rejection were automatically performed prior to download from the Mikulski Archive for Space Telescopes.
Additionally, one-dimensional (1D) spectra were automatically extracted, wavelength calibrated, and flux calibrated, prior to download for all G230LB and G430L observations except the G230LB observation on day 66. 
At this phase, the UV flux was very low at the blue end, preventing the automatic identification of the spectrum location by the pipeline. 
These observations were manually extracted with the \texttt{stistools.calstis} pipeline, using the red end of the spectrum to identify the location of the trace.

The G750L grating suffers fringing at the reddest wavelengths, which can be corrected using a contemporaneous fringe flat. 
We obtained a fringe flat with each visit using the $0.\arcsec3 \times 0.\arcsec09$ aperture, as recommended for the E1 aperture position.
The fringe flat was applied to the G750L observations using the module \texttt{stistools.defringe} and 1D spectra were extracted using the \texttt{stistools.x1d} routine.

We utilized data-quality flags 16 (high dark rate) and 512 (bad reference pixel) to eliminate bad pixels. 
In particular, we note that narrow emission features that appear in all spectra are flagged as high dark rate pixels and therefore removed from the analysis.
We confirm that these are isolated pixels with high count rates in the dark files by examining the extraction location in the two-dimensional dark images themselves. 
Nevertheless, it is possible that there is a real unresolved emission line at the same pixel as at least one of these features.
However, this would be contaminated by the high dark rate and therefore unable to provide any constraint on the CSM. 
Thus, given the characteristics described above, we treat these features as artifacts and do not consider them further.

\begin{deluxetable*}{cccccccc}
\tablecaption{{\it HST}/STIS observations of SN~2023ixf from GO-17313\label{tab:HSTspec}}
\tablehead{\colhead{Phase (d)} & \colhead{JD} & \colhead{Grating} & \colhead{Exposure Time (s)} &  \colhead{Resolving Power} & \colhead{Wavelength Range (\AA)} & \colhead{HST File Rootname}}
\startdata
14.483 & 2460097.733 & G230LB & 1158   & 700 & 1680--3060 & of4301010 \\
14.502 & 2460097.752 & G430L  & 36     & 500 & 2900--5700 & of4301020 \\
14.506 & 2460097.756 & G750L  & 28     & 500 & 5240--10270 & of4301030 \\
19.960 & 2460103.210 & G230LB & 1244   & 700 & 1680--3060 & of4302010 \\
19.979 & 2460103.229 & G430L  & 32     & 500 & 2900--5700 & of4302020 \\
19.983 & 2460103.233 & G750L  & 24     & 500 & 5240--10270 & of4302030 \\
24.722 & 2460107.972 & G230LB & 1232   & 700 & 1680--3060 & of4303010 \\
24.740 & 2460107.990 & G430L  & 34     & 500 & 2900--5700 & of4303020 \\
24.744 & 2460107.994 & G750L  & 26     & 500 & 5240--10270 & of4303030 \\
49.885 & 2460133.135 & G230LB & 3.1     & 700 & 1680--3060 & of4304010 \\
49.903 & 2460133.153 & G430L  & 1.5      & 500 & 2900--5700 & of4304020 \\
49.907 & 2460133.157 & G750L  & 1.5      & 500 & 5240--10270 & of4304030 \\
66.514 & 2460149.764 & G230LB & 1236   & 700 & 1680--3060 & of4305010 \\
66.533 & 2460149.783 & G430L  & 32     & 500 & 2900--5700 & of4305020 \\
66.536 & 2460149.786 & G750L  & 24     & 500 & 5240--10270 & of4305030 
\enddata
\end{deluxetable*}
We add to this dataset the {\it HST} observations taken between days 3 and 11 as part of GO-17205 \citep{2022Zimmerman}, which have been presented by \citet{2023Zimmerman}. 
1D-extracted and flux-calibrated files were downloaded from the Mikulski Archive for Space Telescopes.
As noted by \citet{2023Zimmerman}, some of these exposures suffered from pointing, acquisition, and/or saturation issues, which led to unreliable flux due to the supernova not being aligned in the slit and the nonlinearity of the CCD response near the saturation point. 
To mitigate these issues, we scale all observations with each grating in a given visit to the spectrum with the most flux (assuming this is the best-centered observation) using a first-degree polynomial fit to the ratio of the spectra as the scale factor.
A constant multiple of a few percent was then applied to align the G430L and G750L observations to the G230LB observation, and then the fluxes were combined by taking the median value at each wavelength, excluding pixels with data-quality flags of 16 and 512. 
We note that the majority of pixels are saturated in the range $\sim 3200$--5000\,\AA\ in visit 1, $\sim 2100$--7600\,\AA\ in visits 2 and 3, and $\sim 2900$--7900\,\AA\ in visit 4, making these wavelength ranges the most unreliable in terms of flux calibration. 
However, in this paper, we are primarily analyzing the later epochs ($14-66$ days) which are not saturated and only use the early epochs ($<10$ days) to identify the presence and timing of relevant spectral features.
Thus, we do not attempt to further correct for this effect.

\section{Spectral Evolution} \label{sec:SpecEvolve}
\begin{figure*}
    \centering
    \includegraphics[width=\textwidth]{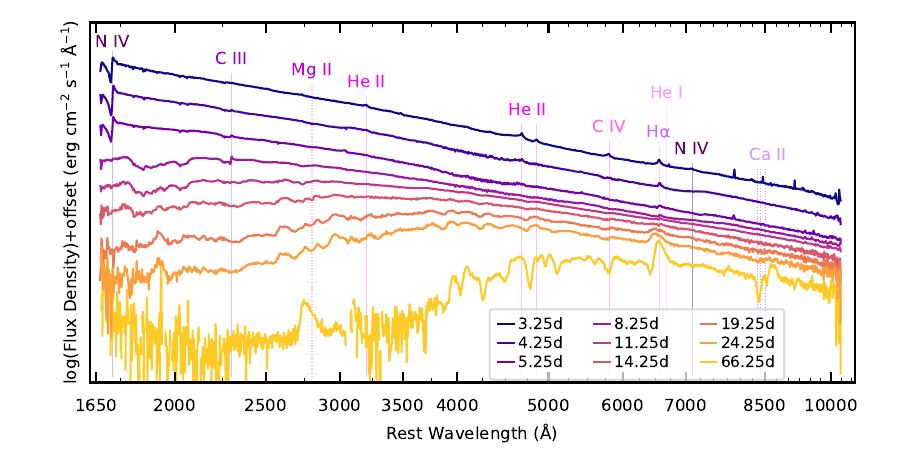}
    \caption{Evolution of the NUV to NIR spectra of SN~2023ixf from 3.5 to 66 days after explosion. Prominent features (\ion{N}{4}: 1718, 7109, 7122 \AA; \ion{C}{3}: 2297 \AA; \ion{Mg}{2}: 2796, 2802 \AA; \ion{He}{2}: 3203, 4685.5, 4860 \AA; \ion{C}{4}: 5801, 5812 \AA; \Ha: 6563 \AA; \ion{He}{1}: 6678.1 \AA; \ion{Ca}{2}: 8498, 8542, 8662 \AA) are marked with vertical lines at their rest wavelength and labeled at the top of the figure. Narrow ISM absorption lines have been removed to highlight the supernova spectra. }
    \label{fig:timeseries}
\end{figure*}

The near-UV (NUV) through NIR evolution of SN~2023ixf from day 3 to 66 is shown in \autoref{fig:timeseries}.  
No Type II supernovae (excluding Type IIn supernovae) since SN~1987A have been observed this early in the UV and with this cadence, revealing unprecedented details of the evolution.
As has been noted by other authors, narrow emission lines are present in the early-time spectra, including notably a narrow P~Cygni line from \ion{N}{4} $\lambda 1718$ which is present in the first spectrum and fades by day 8, and \ion{C}{3} $\lambda2297$ which is present from day 4 to day 8 \citep{2023Yamanaka, 2023Jacobson-Galan, 2023Bostroem2, 2023Teja, 2023Hiramatsu, 2023Zimmerman}.  
These early narrow lines indicate dense CSM surrounding the supernova that has not been shocked.
By day 14, these lines disappear and the optical spectrum is that of a young Type II supernova. 
Although slow to develop as a result of early interaction, a typical Type II supernova optical spectrum emerges with prominent Doppler-broadened P Cygni profiles in \ion{H}{1}, \ion{He}{1}, and other metals.

\begin{figure*}
    \centering
    \includegraphics[width=\textwidth]{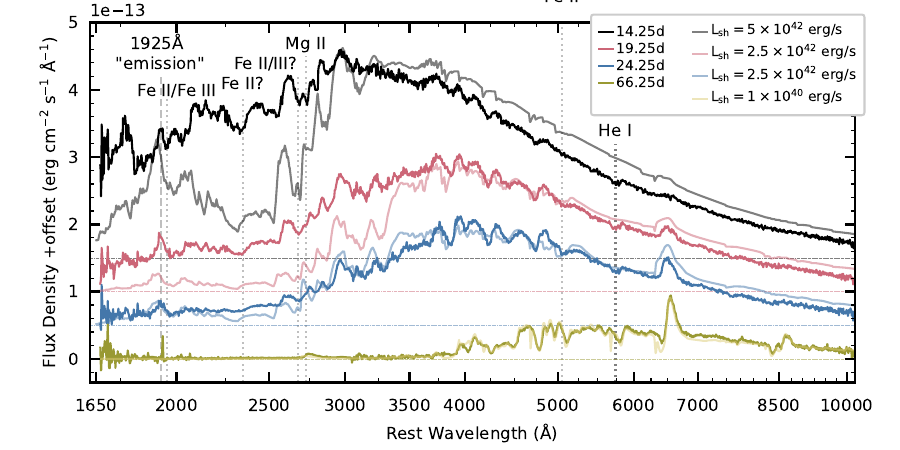}
    \caption{A comparison of the spectra of SN~2023ixf from 14.25 to 66.25 days post-explosion (solid lines) with the best-matching CSM interaction model (semi-transparent). 
    The zero flux levels for each epoch are drawn as a dot-dashed line in the same color as the observed spectrum at that epoch.
    Over time, the UV emission declines and P~Cygni features develop in the optical, until day 66, when only \MgIIlambda\ is visible in the UV. 
    This corresponds to a sequence of models with decreasing interaction power. 
    Even though the model spectra do not match the SED of the observations, especially in the first two spectra where they appear to be cooler (resulting in lower-ionization species and more UV line blanketing), the match is impressive given the 1D nature and the ad-hoc implementation of the mass, density, and clumping of the cool dense shell. 
    Prominent spectral features have been marked with dotted lines at 7000 \kms, the approximate location of the absorption throughout the spectroscopic sequence.
    The ``emission'' feature at 1925~\AA\ is marked at rest with a dashed line.
    Narrow ISM absorption lines have been removed to better highlight the supernova spectra.
    Observed and model spectra at each epoch have been placed at the same distance with the same offset applied.
    }
    \label{fig:ModelComp}
\end{figure*}

A more detailed view of the evolution from day 14 through day 66 is shown in \autoref{fig:ModelComp}.
Although still dominated by iron line blanketing (the broad absorption of flux due to a forest of iron lines that blend together), the UV continuum flux is clearly present in the day 14 spectrum. 
This is unusual for classical Type II supernovae in which the UV flux fades rapidly \citep{2000Baron, 2007Brown, 2023Vasylyev, 2023Bostroem1}.

While the flux blueward of 2500\,\AA\ fades significantly by day 20, prominent absorption features can still be seen in the range 2500--3500\,\AA, including \MgII{}, and just blueward \ion{Fe}{2}/\ion{Fe}{3}.
In contrast with the fading UV flux, a prominent ``emission'' line arises around 1925\,\AA\ which reaches peak strength at 25 days before decreasing in strength. 

By day 66, there is very little UV flux as the supernova ejecta cool causing the SED to shift to longer wavelengths and iron absorption to increase in strength. 
However, the \MgIIlambda{} doublet is clearly present as a boxy broad emission line with a slanted top.
Also, at this time \Ha{} has fully developed a broad P~Cygni profile, and other metal lines typical of Type II supernovae are present in the optical, including the \ion{Ca}{2} NIR triplet $\lambda\lambda\lambda$8498, 8542, 8662. 

\begin{figure*}
    \centering
    \includegraphics[width=\textwidth]{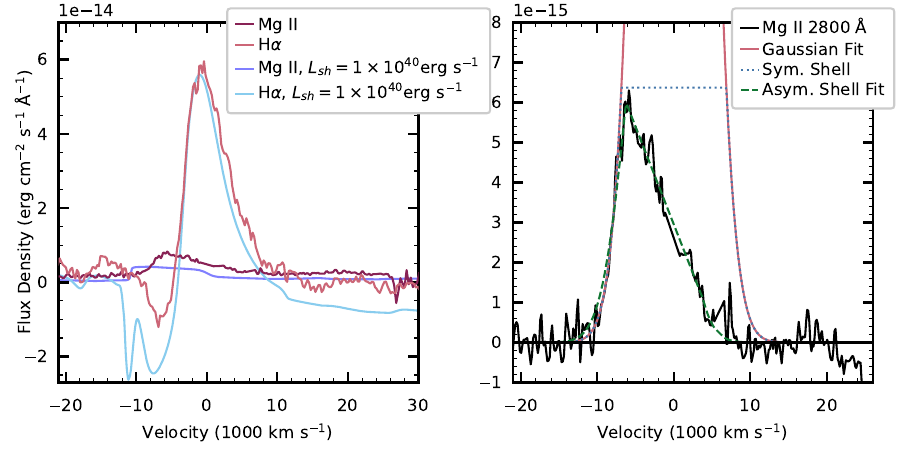}
    \caption{\textit{Left:} \Ha{} and \MgII\ line profiles on day 66 in velocity space of SN~2023ixf compared to the CSM interaction models of \citet{2022Dessart} with $L_{\rm sh}=1\times10^{40}$ \pwrunit{}, scaled to the distance of SN~2023ixf. 
    The shapes of both lines are well captured by the models, given that the cool dense shell is at a higher velocity in the models than in SN~2023ixf.
    \textit{Right:} \MgII\ profile (black) fit with a pure Gaussian (pink) modeling spherical ejecta, a symmetric shell (blue dotted) modeling emission from the cool dense shell in the absence of optical-depth effects, and an asymmetric shell (green dashed) modeling the cool dense shell with optical depth. }
    \label{fig:MgII_Ha}
\end{figure*}

A comparison of the \MgII\ and \Ha\ profiles provides insight into the origin of the \ion{Mg}{2} emission (see \autoref{fig:MgII_Ha}). 
When CSM is present, a cool dense shell forms at the interface between the ejecta and CSM. 
While the \Ha\ emission originates from the ejecta and the cool dense shell, the \MgII\ is thought to originate only from the shell and should thus be at a similar velocity as the outer edges of the ejecta.
We compare \MgII{} and \Ha{} in \autoref{fig:MgII_Ha}, finding that the profiles extend to similar velocities on both the red and blue sides, confirming that the \MgII\ is emitted from the cool dense shell at the outer edge of the ejecta. 
However, \Ha\ sits in the optical where there is continuum and forms throughout the ejecta and cool dense shell, leading to a P Cygni profile, while \MgII{},  without UV continuum and forming in the outer ejecta, is seen solely in emission.
There is a shallow, narrow dip at the blue edge of the \Ha\ profile that could be absorption from the cool dense shell at $\sim9200$ \kms, although it is not clear if this feature is real given the signal-to-noise ratio (S/N) of the spectrum. 

To characterize the \MgII{} feature further, we fit three different profiles in \autoref{fig:MgII_Ha}, representing different physical configurations.
A spherical distribution of material with a Gaussian emitting profile will produce a Gaussian profile.
Although the profile is clearly not Gaussian, it could be coming from a spherical distribution of material that is being attenuated.
To explore the nonattenuated spherical distribution, we fit a Gaussian to the blue side of the \MgII{} profile from $-10,000$ to $-7000$ \kms. 
This fit is shown as a pink line in \autoref{fig:MgII_Ha}; it has a mean of $\mu=0$ \kms\ and a FWHM of 7780 \kms. 
The observed blue side of \MgII{} is well described by a Gaussian profile. 
An alternate emitting region could be a sphere with a Gaussian profile that has had the center removed to form a thick shell.
This configuration will have a flat-topped profile with Gaussian wings \citep{2017Jerkstrand} if the cool dense shell and ejecta are optically thin.
Given our slanted top profile, we use the maximum of our observed flux as the flat-topped maximum of the model profile.  
This defines the inner radius velocity of the shell as $v_{\rm in}= -6800$ \kms, and the profile is shown as a dotted-blue line in the right panel of \autoref{fig:MgII_Ha}.

To model the slated top of the profile, we model the line as if it were generated by an asymmetric shell of material (i.e., a shell in which there is more flux from one side than the other reaching the observer).
Following \citet{2023Kwok, 2024Kwok}, we model this asymmetric shell as a Gaussian profile with an off-center hole cut out, which results in a slanted top instead of a flat-topped profile.
We parameterize this model by the FWHM and mean of the Gaussian, the velocity of the center of the hole ($v_{\rm c}$), and the velocity of the radius of the hole ($v_{\rm in}$). 
We find  FWHM = 7580 \kms, $\mu=-2995$ \kms, $v_{\rm in}=5000$ \kms, and $v_{\rm c}=1750$ \kms, where $v_{\rm c}$ and $v_{\rm in}$ are measured relative to $\mu$. 
This model fits the data exceptionally well and is shown as a green dashed line in the right panel of \autoref{fig:MgII_Ha}.
While the CSM is likely asymmetric, we caution that this parameterization is more likely to indicate different optical depths along the line of sight through the ejecta and cool dense shell, and should not be taken as a literal description of the ejecta.
It does, however, demonstrate that this profile can be created by a shell of material in which a higher flux of photons from the blue side of the ejecta is reaching us, than photons from the red side.
Additionally, it defines the inner edge of the blue side of the shell to be at velocity $-6245$ \kms (where the profile peaks), close to our earlier estimate of $-6800$ \kms. 
From these two values, we adopt $v_{\rm in}$ = -6500 \kms. 
From these fits we conclude that the emitting region of the \MgII\ is well described by a Gaussian profile with a FWHM $\approx$ 7700 \kms\ and an inner radius of $v_{\rm in}$ = -6500 \kms\ that has been attenuated on the red side due to the opacity of the ejecta and cool dense shell. 
We note that as the profile approaches the peak flux from the blue side, the rise is shallower than the Gaussian profile, possibly indicating asymmetry.
Finally, we note that if the CSM is asymmetric, inner and outer shell velocities would represent projected velocities.

Narrow absorption lines are present in the spectra from \ion{Fe}{2} $\lambda2344$, $\lambda2374$, $\lambda2383$, $\lambda2587$, $\lambda2600$ and \ion{Mg}{2} $\lambda\lambda2796,2802$, $\lambda2852.96$.
These lines are centered at the host redshift and there is no evolution in their equivalent widths with time. 
From this, we conclude that these lines are either from the interstellar medium (ISM) in the host galaxy or distant CSM and remove them from the remainder of the analysis.

\section{Comparison with Literature Models} \label{sec:CompModel}
Although by day 14 the narrow emission lines typically associated with strong CSM interaction have faded, it is possible that some energy from the interaction of the supernova ejecta with CSM is still contributing to the observed supernova spectrum. 
When supernova ejecta encounter CSM, a cool dense shell is formed between the forward and reverse shocks. 
The cool dense shell absorbs and reemits some of the shock power from CSM interaction. 
\citet{2022Dessart} published 1D (spherically symmetric) models of the interaction of supernova ejecta and cool dense shell with a low-density wind ($\rm \dot{M}<10^{-3}$ \masslossunit).
Unlike those of \citet{2017Dessart}, which model interaction in a spherically symmetric dense, confined CSM, these models ignore the hydrodynamics of the CSM interaction in exchange for more accurate radiative transfer.
While \citet{2022Dessart} compared their models to optical observations, they show that the signatures of this interaction are strongest in the UV, and our spectral series allows us to test these model predictions. 

Briefly, these models start with a 15 \msun{} progenitor of solar metallicity evolved with Modules for Experiments in Stellar Astrophysics \citep{2011Paxton,2013Paxton} which has been exploded with the radiation-hydrodynamics code \texttt{V1D} \citep{1993Livne,2010Dessart1} and evolved to 10 days post explosion.
On day 10, these observations are mapped to the nonlocal thermodynamic equilibrium radiation transport code CMFGEN \citep{1998Hillier, 2012Hillier, 2013Dessart, 2019Hillier} assuming homologous expansion after this point.
A cool dense shell is simulated by placing 0.1 \msun{} of material at 11,700 \kms\ into the density profile. 
This velocity was chosen as the shock and cool dense shell are not expected to slow down much in the presence of a low-density CSM over a time scale of weeks \citep{2017Dessart}.
The power from the CSM-ejecta interaction is then deposited at a constant rate into this region, mimicking the conversion of kinetic energy from the ejecta into radiative energy. 
Implicit in this implementation is the assumption that the cool dense shell is fully formed during the first few days after the explosion and does not grow in mass.

With this setup, they produce models with no power deposited and continuous depositions of $1\times10^{40}$, $1\times10^{41}$, $5\times10^{41}$, $1\times10^{42}$, $2.5\times10^{42}$, $5\times10^{42}$, and $1\times10^{43}$ \pwrunit\ from day 15 through at least day 120 (some models are run longer).
We emphasize that these models were not created for this specific supernova (or any supernova with early narrow emission lines). 
One difference between the model setup and SN~2023ixf is the presence of confined dense CSM, which produces the flash features observed in the early spectra.
\citet{2017Dessart} find that in the presence of dense CSM, the maximum ejecta velocity decreased from 11,500 \kms\ in the low-density scenario to 7200 \kms\ in the high-density scenario \citep[see also ][]{2023Dessart}. 
We thus expect the velocity of the outer ejecta of SN~2023ixf to be slower than the models of \citet{2022Dessart}. 
Another consequence of the dense CSM is that the initial conditions of the interaction region of the model may not be representative of the physical conditions of the outer ejecta of SN~2023ixf, leading to a larger mismatch between the observations and the models in the earliest epoch than at later times. 

We visually compare each model at a given epoch to our observed flux, starting from the day 14.25 spectrum, and identify the model that best matches the observed flux.
Although each of the models represents a steady-state wind, the cool, dense shell is optically thin to continuum radiation and radiates efficiently.
For this reason, it will react promptly to a change in shock power. 
Thus, we expect that a decreasing mass-loss rate with radius would effectively have the same result as taking a different steady-state model for different epochs.
The spectra of SN~2023ixf and best-matched models are shown in \autoref{fig:ModelComp} and the model properties are listed in \autoref{tab:CSMParams}.
As expected, the model velocity is faster than the observed ejecta velocity leading to blueshifted and broader features. 
We find that the observations of SN~2023ixf indicate a progressive drop in interaction power over time, which translates into a drop in progenitor mass loss with increasing distance.

\begin{deluxetable*}{cccccc}
\tablecaption{Characteristics of the CSM around the progenitor of SN~2023ixf assuming $v_{\mathrm{shock}}=10000$ \kms\ and $v_{\mathrm{wind}}=55$ \kms. \label{tab:CSMParams}}
\tablehead{\colhead{Phase} & \colhead{Shock Luminosity} & \colhead{Density} &\colhead{Mass-loss Rate} & \colhead{Distance} & \colhead{Time Before Explosion } \\ 
\colhead{ (yr)} & \colhead{(\pwrunit)} & \colhead{(\densityunit)}  & \colhead{(\masslossunit)} & \colhead{(cm)} & \colhead{(yr)} } 
\startdata
       14.25  & $5\times10^{42}$   &$5.2\times10^{-16}$ &   $8.7\times10^{-4}$   &  $1.2\times10^{15}$ & 7 \\
       19.25  & $2.5\times10^{42}$ &$1.4\times10^{-16}$ &   $4.4\times10^{-4}$ &  $1.7\times10^{15}$ & 10 \\
       24.25  & $2.5\times10^{42}$ &$9.1\times10^{-17}$    &   $4.4\times10^{-5}$ &  $2.1\times10^{15}$ & 12 \\
       66.25  & $1\times10^{40}$   &$4.8\times10^{-20}$    &   $1.7\times10^{-6}$ &  $5.7\times10^{15}$ & 33  \\
       66.25  & $1\times10^{41}$   &$4.8\times10^{-19}$    &   $1.7\times10^{-5}$ &  $5.7\times10^{15}$ & 33 
\enddata
\tablecomments{The last two rows bracket the true shock luminosity as the \MgIIlambda\ emission lies between $L\mathrm{_{sh}=10^{40}}$ and $L\mathrm{_{sh}=10^{41}}$ \pwrunit.}
\end{deluxetable*}
Considering that these models were not produced for this specific supernova, it is impressive that no scaling beyond correcting for distance is required to match the observations and many of the features are reproduced, although at varying strengths.
The first two epochs of models are too faint in the NUV and too bright in the optical, a discrepancy that improves with time.
Nevertheless, we consider the best-matched model to be indicative of the CSM interaction level as the selected model most closely matches the observed flux: the model with more CSM interaction is too bright at all wavelengths and the model with less CSM interaction is even more discrepant for the majority of the spectrum. 
This consideration is further validated by the custom modeling in \autoref{sec:discussion}.
One possible reason for the difference between the models and observations is that the temperature is too low in the emitting region in the early models. 
This is corroborated by the lack of \ion{He}{1} $\lambda$5875 in the models, whose formation requires higher temperatures  \citep{2005Dessart}

\begin{figure*}
    \centering
    \includegraphics[width=\textwidth]{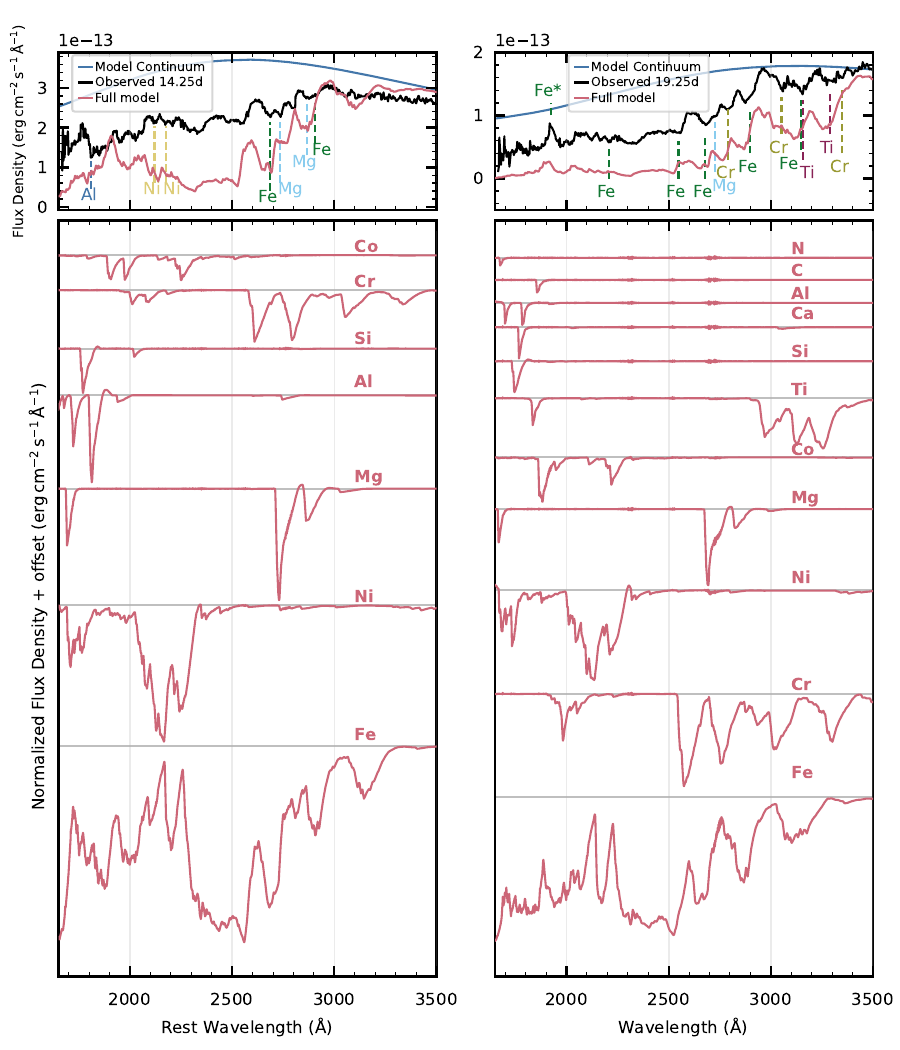}
    \caption{\textit{Top:} UV spectra of SN~2023ixf on day 14.25 (left) and 19.25 (right) compared to radiative-transfer models. Absorption lines from individual species are marked below the spectra. 
    In the day 19.25 spectrum, the asterisk on Fe at 1924 \AA\ indicates that this is a window of minimal iron absorption rather than an emission line due to any atomic transition.
    The individual features are reproduced quite well, albeit at higher velocities in the models.
    \textit{Bottom:} decomposition of the model spectrum at each epoch into species that contribute to the spectrum. 
    Although the individual spectra are offset, the scales of the spectra are consistent, showing the relative intensity of each species.
    The dominance of iron-group elements (and in particular iron) in the spectra at both epochs is clear.
    }
    \label{fig:Ladder}
\end{figure*}

The UV is dominated by a forest of iron-group-element absorption lines that blend together, making it challenging to associate features with the emission or P~Cygni profiles of individual species \citep{2023Bostroem1,2005Dessart}. 
To facilitate the identification of regions with contributions from different elements, we post-process the CMFGEN models to calculate the spectra on day 14 and day 19, omitting the bound-bound transitions of individual species. 
The full spectrum is then divided by these spectra to obtain the relative flux contribution of each species. 
We perform this calculation for H, He, C, N, O, Ne, Na, Mg, Al, Si, S, K, Ca, Sc, Ti, Cr, Fe, Co, and Ni.
Comparing the omitted spectrum to the full spectrum for each species, regions were identified as at least partially resulting from an individual species. 
These are marked in the top panel of each column of \autoref{fig:Ladder}, which shows the full model spectrum compared to the observed spectrum at the same epoch.
Elements that have absorption features with a depth of $>2$\% of the total flux are shown in the bottom panel of \autoref{fig:Ladder}.
In these panels, the strong influence of the iron-group elements (specifically Fe and Ni on day 14 and Fe, Ni, and Cr on day 19) on the overall spectrum is clear. 
In particular, we point out that the apparent developing ``emission line'' at 1924\,\AA\ is in fact a window of lower absorption between two prominent iron absorption regions, similar to the ``emission line'' at 2970\,\AA\ identified in the NUV spectra of SN~2022acko on day $\sim 20$ \citep{2023Bostroem1}.  
Additionally, we show the continuum emission if no absorption is present in the top panel of each column of \autoref{fig:Ladder}, demonstrating how strongly the spectrum deviates from blackbody emission.

\begin{figure}
    \centering
    \includegraphics[width=\columnwidth]{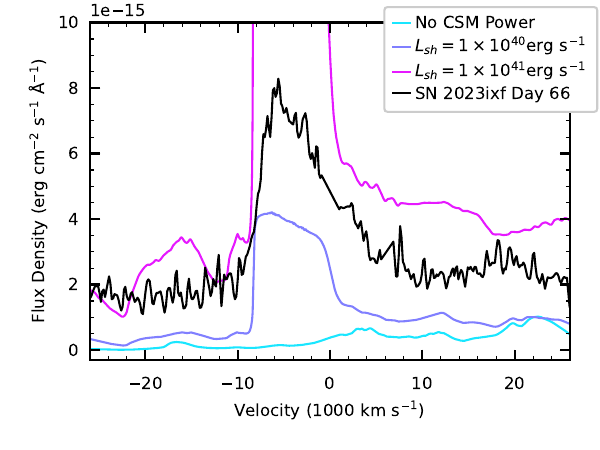}
    \caption{\MgII\ line profiles of SN~2023ixf on day 66 in velocity space (black) compared to models with no power from CSM interaction (cyan) and the two lowest levels of CSM interaction modeled ($L_{\rm sh}=1\times10^{40}$ \pwrunit; blue, $L_{\rm sh}=1\times10^{41}$ \pwrunit; magenta). 
    This feature is present only in those models with CSM interaction, indicating that SN~2023ixf is interacting with CSM on day 66.
    The velocity scale of the models has been scaled to match the observations.}
    \label{fig:MgII_models}
\end{figure}

Another emission line of particular note is  \MgII\ in the day 66 spectrum, which is compared to the CMFGEN models in \autoref{fig:MgII_models}. 
To account for the higher velocity of the models, we scale the models by the ratio of the blue edge of the feature in the observation to the blue edge in the model which preserves the morphology of the line. 
The \MgII\ feature is only present in models with CSM interaction after day $\sim 50$ and rapidly increases in strength for greater interaction power.
Our observation on day 66 falls between the models with a shock power of $L_{\rm sh}=1\times10^{40}$~\pwrunit{} and $L_{\rm sh} = 1\times10^{41}$~\pwrunit{} and looks clearly distinct from models with no interaction.

In the left panel of \autoref{fig:MgII_Ha}, we compare the \MgII\ and \Ha{} profiles of our observation and the $L_{\rm sh}=1\times10^{40}$ \pwrunit\ model (without any shift in wavelength) on day 66.
We find that both profiles are similar, although the model extends to bluer wavelengths due to the lack of dense CSM in the early supernova evolution, which caused the cool dense shell to form at lower velocities in SN~2023ixf.
While the models have a large narrow absorption feature on the blue side of \Ha\ from the blue side of the cool dense shell, no such prominent feature is present in the observed \Ha\ profile. 
There is a hint of an absorption feature around 9200 \kms, although this is near the noise threshold of the spectrum and thus we cannot confidently associate it with this feature. 
On the red edge of \Ha{}, the models show a subtle red shelf.
With the S/N of our data, we cannot verify whether this feature is present.

The model \MgII{} is very boxy, rising sharply at the blue edge of the \Ha{} profile to a plateau before falling off just after 0 \kms. 
This shape arises from a shell of material with the back side obscured by the cool dense shell and ejecta, leading to what appears to be blueshifted emission.
Compared to our observations, the \MgII\ feature in the models is boxier and more blueshifted. 
Additionally, the \MgII{} emission in the observations rises more gradually to a peak and begins to fall off immediately, rather than plateau. 
Nevertheless, the observations have the same basic structure as the models, indicating a similar emitting mechanism and structure.

\section{Comparison with Other Supernovae}\label{sec:CompObs}
We focus our comparison of SN~2023ixf to other supernovae on the small number of Type II supernovae with UV observations, defining a sample of objects with a range of CSM properties and supernova types (see \autoref{tab:compSN}). 
Although many of these objects continued to exhibit signs of CSM interaction later in their evolution \citep[e.g.,][]{2002Pooley, 2012Mauerhan, 2017Smith2}, we limit the discussion to their plateau-phase evolution, which corresponds to our observations of SN~2023ixf. 

SN~1993J was a Type IIb supernova, with the hydrogen lines fading and helium lines increasing in strength soon after explosion \citep{1993Filippenko}. 
Many lines of evidence indicated interaction with CSM including narrow emission lines \citep{1994Benetti} and X-ray \citep{1994Zimmermann} and radio \citep{1993Pooley} detections. 
SN~1979C and SN~1998S are both Type II supernovae with steeply declining light curves and above-average photospheric phase luminosities. 
They also showed narrow emission lines for weeks after explosion; however, these did eventually disappear \citep{1980Panagia, 2000Leonard}. 
SN~1979C was also detected at X-ray and radio wavelengths indicating interaction with CSM beyond the nominal RSG wind \citep{1980Panagia}. 
SN~1980K had a decline rate similar to that of SN~1979C, although it was about a magnitude fainter in the $V$ band. 
While it did not exhibit narrow emission lines, it was detected in X-rays \citep{1982Canizares} and radio \citep{1992Weiler}.
SN~1987A was a peculiar Type II supernova, originating from a compact, blue supergiant progenitor. 
No CSM interaction was initially detected; however, months after explosion narrow emission lines appeared as high-energy photons ionized a ring of CSM. 
SN~1999em, 
SN~2021yja, and SN~2022wsp are considered fairly normal Type II supernovae \citep{2000Leonard, 2000Baron, 2008Gal-Yam, 2022Hosseinzadeh, 2022Vasylyev, 2023Vasylyev}, although \citet{2022Hosseinzadeh} note that SN~2021yja was exceptionally blue at early times and \citet{2023Vasylyev2} find that the absorption of \Ha{} and \Hb{} in SN~2022wsp was suppressed in the P~Cygni profiles which they ascribe to CSM interaction. 
Finally, SN~2005cs and SN~2022acko were low-luminosity supernovae \citep{2007Brown, 2023Bostroem1}.
The only possible sign of CSM interaction in these supernovae is a ``ledge'' feature in the early spectra, which has been attributed to CSM, although other possibilities exist \citep[see ][for a detailed discussion]{2023Pearson}. 
SN~2021yja also showed the `ledge' feature.
Details on the explosion epochs, extinction values, and spectra considered in this analysis are given in \autoref{tab:compSN}.
We note that UV data also exist for SN~2010jl, a superluminous Type IIn supernova.
Throughout its evolution, the UV spectrum is dominated by the cool dense shell with narrow spectral features and enhanced flux, in contrast to the other supernovae in this sample where the spectra are dominated by the supernova photosphere. 
We therefore exclude it from our analysis. 

\begin{figure*}[h]
    \centering
    \includegraphics[width=\textwidth]{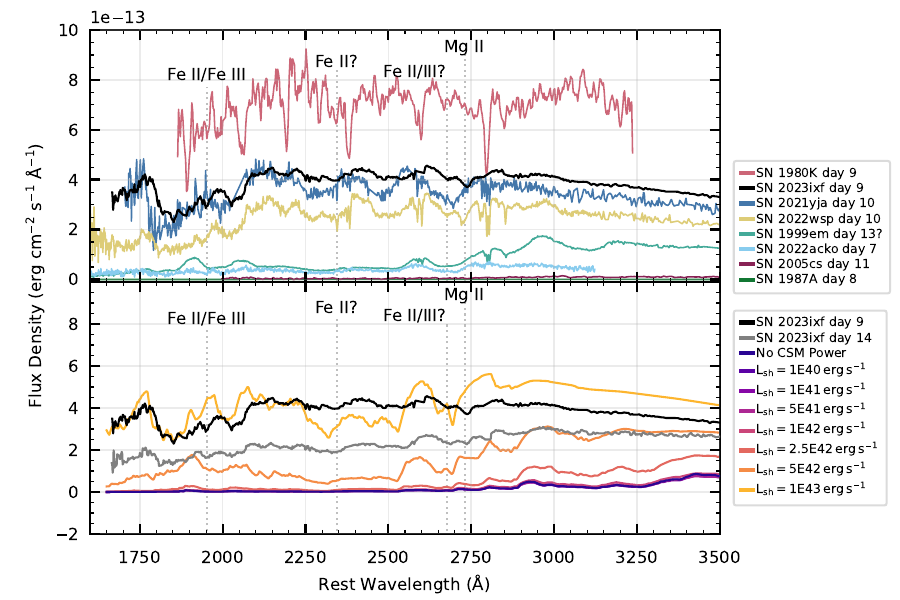}
    \caption{\textit{Top}: comparison of the UV spectrum of SN~2023ixf on day 9 with that of other Type II supernovae at a similar epoch (phase for each supernova given in the panel legend; all spectra scaled to the distance of SN~2023ixf): SN~1980K, SN~1987A, SN~1999em, SN~2005cs, SN~2021yja, SN~2022wsp, and SN~2022acko. 
    At NUV wavelengths, SN~2023ixf is among the brightest in the sample. 
    \textit{Bottom}: day 9 (black) and 14 (gray) spectra of SN~2023ixf compared with CMFGEN models having varying degrees of shock power on day 14. 
    Although the majority of UV observations have phases around day 10, day 14 is the first epoch modeled by \citet{2022Dessart}. 
    We show both SN~2023ixf spectra to demonstrate possible evolution from the spectra in the top panel and the models in the bottom. The day 14 spectrum best matches the model with $L_{\rm sh}=5\times10^{42}$ \pwrunit{} below $\sim 3000$\,\AA{} but is brighter than the model at shorter wavelengths.
    Elements are marked in both panels in black at $-7000$ \kms\ to match the SN~2023ixf absorption. }
    \label{fig:ObsComp10}
\end{figure*}

\begin{figure*}[h]
    \centering
    \includegraphics[width=\textwidth]{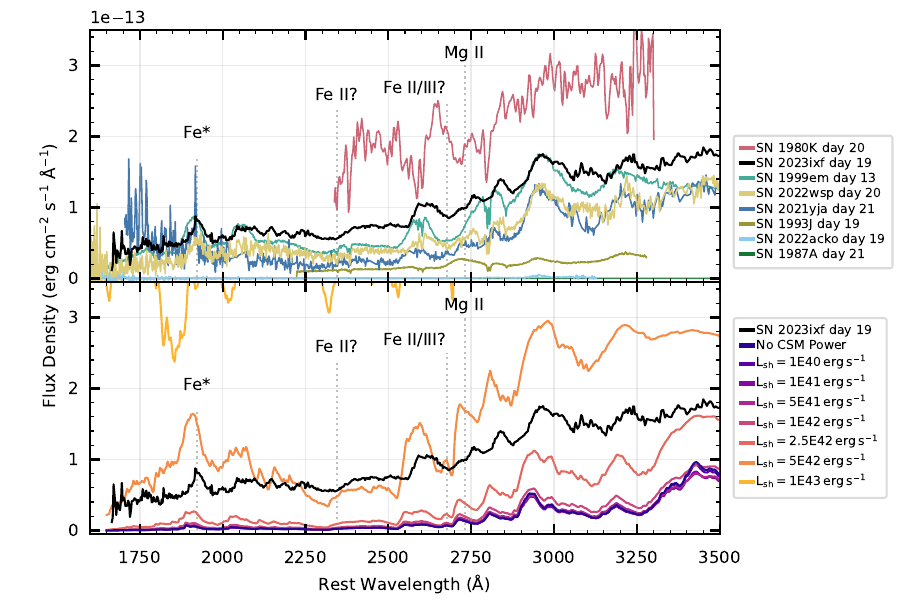}
    \caption{\textit{Top}: comparison of the spectrum of SN~2023ixf on day 19 with that of other Type II supernovae at a comparable epoch (phase for each spectrum given in the panel legend; all spectra scaled to the distance of SN~2023ixf): SN~1980K, SN~1987A, SN~1993J,  SN~2021yja, SN~2022wsp, and SN~2022acko. SN~2023ixf remains UV bright, especially when compared with SN~2021yja, which has faded to a flux similar to that of SN~2022wsp. 
    The explosion epoch of SN~1999em is uncertain and thus we show the spectrum here, too, to demonstrate its similarity to the other spectra at these epochs.
    \textit{Bottom}: UV spectrum of SN~2023ixf on 19 days (black) compared with the CMFGEN models having varying degrees of shock power on day 20 (colors). Although at optical wavelengths the spectrum best matches the model with $L_{\rm sh}=2.5\times10^{42}$, below $\sim 2750$\,\AA{} the spectrum best matches the model with $L_{\rm sh}=5\times10^{42}$. 
    Elements are marked in both panels in black at $-7000$ \kms\ to match the SN~2023ixf absorption.
    The asterisk denotes the feature caused by a window between two strong iron absorption troughs.}
    \label{fig:ObsComp20}
\end{figure*}

In \autoref{fig:ObsComp10} and \autoref{fig:ObsComp20} we show a comparison of the spectra of SN~2023ixf to this sample of UV spectra, scaled to the distance of SN~2023ixf, on day $\sim 10$ and day $\sim 20$, respectively.
For this comparison, we exclude SN~1979C and SN~1998S,  as their UV flux is significantly higher than that of SN~2023ixf and all models considered in this paper.
On day 10, SN~2023ixf is similar in brightness to SN~2021yja. 
Although it has a lower flux, SN~2022wsp is spectroscopically similar. 
The \MgIIlambda\ doublet is present in absorption in SN~1980K, SN~2021yja, SN~2022wsp, and SN~2022acko, although it is blueshifted to higher velocities in SN~2021yja and SN~2022acko.
Additionally, blueward of \MgII, the Fe absorption feature identified by \citet{2023Vasylyev} in SN~2022wsp is clearly present in SN~2023ixf and SN~1980K. 
Although not a pronounced absorption trough, it is also possibly present in SN~2021yja and SN~2022acko and blended with \MgII{} in SN~1999em.
Like SN~2023ixf, the UV spectra of SN~1980K, SN~2021yja, and SN~2022wsp all show an absorption complex of Fe and Ni between 1750\,\AA{} and 2050\,\AA. 
The Fe absorption complex between 2050\,\AA{} and 2550\,\AA, which is notably shallower in SN~2023ixf and SN~1980K than models predict, is fairly well reproduced in SN~2021yja, although all of the observed profiles have less prominent individual troughs than the model. 

One of the most distinctive features of CSM interaction is excess UV flux.
On day 10, we see that SN~1980K has a higher flux than SN~2023ixf, SN~2021yja has a very similar flux while SN~2022wsp is a bit lower, SN~2022acko is significantly lower, and the remaining supernovae have virtually no flux on day 10. 
We note, however, that both excess UV flux and the slope of the UV SED are highly dependent on the assumed distance, extinction, and explosion epoch of the supernovae. 
In particular, the redshift-independent distance of NGC 4269 ranges from 4 to 7.8 Mpc \citep[see ][for a thorough discussion]{2019VanDyk},  significantly affecting the luminosity of SN~1980K. 
While the distances and thus overall luminosities are uncertain, individual features can indicate different levels of CSM interaction.
For example, the development of an Fe absorption complex redward of 3000\,\AA{} is not present in SN~1980K, SN~2021yja, and SN~2023ixf.
This feature only appears in models with lower levels of CSM interaction. 
This line does appear to be developing in SN~2022wsp, consistent with the interpretation that the lower flux level is indicative of less CSM interaction. 
Comparing with the models, this suggests that SN~2021yja and SN~2022wsp have close to $1\times10^{43}$ \pwrunit{} interaction power added to the ejecta luminosity on day 10.

\citet{2023Bostroem1} noted that SN~1999em has an extremely low UV flux relative to comparable sequences. 
It is interesting to consider both the anomalous SED and features of SN~1999em relative to the other supernovae in this sample, as SN~1999em is a relatively prototypical Type IIP supernova \citep[although a high-velocity feature was detected in both \Ha{} and \Hb{} during the photospheric phase, interpreted as a sign of CSM interaction;][]{2007Chugai}. 
The explosion epoch used is halfway between the last nondetection and first detection. 
It is possible that, at discovery, SN~1999em was up to 5 days older, as suggested by \citet{2006Dessart}, which would make the UV flux less unusual; however, even at a later phase, the shape of the UV flux is different from that of the other supernovae --- either more suppressed below 2750\,\AA{} or with increasing flux redward of that. 
The presence of an``emission'' feature owing to the window of absorption at 1924\,\AA{} further supports a later explosion epoch.
Given this uncertainty, SN~1999em is plotted in both \autoref{fig:ObsComp10} and \autoref{fig:ObsComp20}.

On day 20, the UV flux has decreased in all supernovae. 
As was observed in SN~2022wsp \citep{2023Vasylyev}, the Fe feature blueward of \MgII\ has faded in SN~2023ixf. 
Interestingly, SN~2021yja shows an extended feature that encompasses both the Fe and \MgII\ features, possibly indicating that these are blended. 
The slope and features of SN~2021yja and SN~2022wsp are very similar to the model with $L_{\rm sh} = 2.5\times10^{42}$ \pwrunit{}, although they all have a larger flux.
The slope of SN~2023ixf and SN~1999em is not reproduced by any model, even if an overall flux offset is applied. 
It is possible that models customized for each supernova with more sophisticated physics would provide a better fit.
For example, a compelling time-dependent model for SN~1999em is presented in \citet{2006Dessart}.
The flux of SN~2021yja has fallen to a level similar to that of SN~2022wsp, indicating a lower level of power from CSM interaction. 
The fact that the flux of SN~2021yja fell relative to SN~2023ixf and SN~2022wsp from the flux levels on day 10 gives us confidence that this is not a distance effect.
The window of absorption at 1924\,\AA{} creates a clear ``emission line'' in SN~2021yja, SN~2022wsp, and SN~2023ixf.

In interacting supernovae,  the \MgII\ emission is attributed to emission from the cool dense shell, which has been excited by radiation from the reverse shock \citep{1994Chevalier, 2022Dessart}.
Originating from a shell of material, the emission-line profile is expected to be broad, boxy, and blue shifted \citep{2022Dessart}.
It has been noted in SN~1979C 14 yr after explosion \citep{1999Fesen}, in SN~1993J on day 670 \citep{2005Fransson}, in SN~1995N on day 943 \citep{2002Fransson}, in SN~1998S on days 28--485 \citep{2005Fransson}, and in SN~2010jl on days 34--543 \citep{2014Fransson}.
However, these supernovae represent the more strongly interacting objects in our sample. 
It remains to be seen if this feature is present in supernovae that do not have any period of strong interaction with dense CSM.

We searched all available NUV spectra of our comparison sample of Type II supernovae for the broad, boxy \MgII\ emission that was observed on day 66 in SN~2023ixf. 
As \MgII\ is present in absorption on day 25 and then in emission on day 66, we do not know the exact timing of this transition.
Additionally, this timing is related to the ejecta and CSM properties and thus may not be a universal property of Type II supernovae.
We therefore consider all spectra with phases greater than 30 days. 
One complicating factor is the presence of \MgII\ absorption from both the MW and the host galaxy, which contaminates the supernova emission and can make it challenging to assess the shape of the emission profile.
Whenever possible, we remove this narrow absorption from our data.

We do not detect this feature in SN~1987A, although this is perhaps not surprising given its blue supergiant progenitor and lack of UV flux.
The feature is clearly detected in the remaining supernovae as a broad, boxy, blue-shifted profile.
In SN~1980K, we marginally detect it. 
However, the S/N of the spectra is fairly low and the feature is not as strong as in other supernovae.
In SN~1979C we detect it developing already on day 24, although it could be as old as 32 days if the supernova explosion occurred immediately after the last nondetection. 
From this date, \MgII\ evolves from a blue-shifted, flat-topped profile into one with a slanted profile (higher blue edge) starting around day 60.
Similarly, \MgII\ evolves in SN~1998S from a boxy, blue-shifted, symmetric profile on day 72 to an asymmetric profile on days 237 and 485. 
In SN~1993J, the \MgII\ emission is fairly constant with a boxy, asymmetric profile from day 173 to 669.
This is the earliest epoch of data for SN~1993J after the UV continuum has faded.
While the feature is clearly present as broad emission in SN~1995N and SN~2010jl (days 30--572), the ISM absorption from both the Milky Way and the host galaxy obscure the shape of the profile and we do not include these in our analysis.

\begin{figure}
    \centering
    \includegraphics[width=\columnwidth]{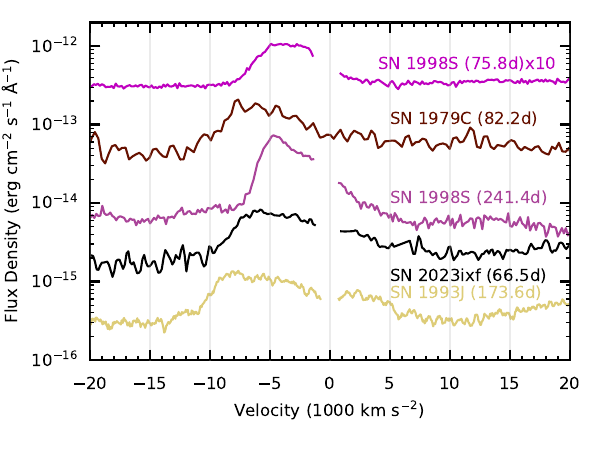}
\caption{Comparison of the \MgII\ emission line in SN~2023ixf (day 66; black) to that of other supernovae that show this feature (SN~1979C, brown; SN~1993J, yellow; and SN~1998S, magenta). All lines have a blueshifted peak. 
In all but the first SN~1998S spectrum, the profiles are asymmetric  indicating that the red side of the line is attenuated.
The gap in some spectra is due to the removal of narrow Galactic and host ISM absorption. }
    \label{fig:MgIIComp}
\end{figure}

The \MgII\ emission features of SN~1979C, SN~1980K, SN~1993J, and SN~1998S are compared to that of SN~2023ixf in \autoref{fig:MgIIComp}. 
We choose the closest available epoch to day 66, and for SN~1998S we also include the first epoch at which the profile is asymmetric.
From these comparisons, we find that if a change in the shape of the \MgII\ emission is observed, the emission begins as a symmetric, boxy, blue-shifted profile and always evolves to the same asymmetric, boxy profile, with more flux on the blue side.
For supernovae that, like SN~2023ixf, always show an asymmetric profile, the asymmetry is always higher on the blue side.
The consistency of this asymmetry tells us that this is not an effect of asymmetric CSM that we would expect to observe from a different viewing angle, but rather an optical-depth effect. 
This is corroborated by the \citet{2022Dessart} models, which show a blue-shifted, boxy emission profile with no emission from ejecta at velocities below 7000 \kms\ (i.e. the emission is isolated to the cool dense shell).
Additional validation is derived from a custom model that reproduces the models of \citet{2022Dessart} but with the cool dense shell at 8500 \kms\ and shows the same slanted profile as we observed in the \MgII\ feature.
While optical-depth effects are responsible for the blue-shifted and slated profile at early times, at later times dust formation likely attenuates the red side of the feature. 
This sample shows a variety in maximum velocity of the blue side of these features, indicating diversity in the location of the cool dense shell, as well as other supernova parameters such as the ejecta mass, hydrogen envelope mass, explosion energy, and CSM mass.
However, they uniformly rise to maximum more slowly than the models, indicating a different cool dense shell density profile than is implemented in the models.

Overall, this comparison confirms that UV spectra are sensitive to different levels of CSM interaction, and the \MgII\ feature can reveal weak CSM interaction even when it is not clearly seen in the optical. 
The sparse time series of the other supernovae that we have assembled indicates a diversity in features and their strength, velocity, and evolution, but a larger sample is required to evaluate further. 
Additionally, while the models reproduce many features, they are not able to reproduce the slope and possibly the overall flux of the observations, indicating room for further detailed modeling.

\section{Discussion}\label{sec:discussion}
\subsection{Testing Model Predictions}
With our sequence of observations, we can test the following predictions of signatures of CSM interaction made by \citet{2022Dessart} through day 66.
\begin{enumerate}
    \item Higher luminosity and flux in all filters for supernovae having more interaction, with this effect being more prominent in the UV and at later times when the ejecta luminosity has faded.
    \item Emergence of the \MgII\ doublet as broad, boxy emission $\sim50$ days after explosion,  increasing in luminosity until day $\sim$175. 
    \item Weakened absorption features as the photospheric spectrum of the supernova is filled in by emission from the cool dense shell. 
    \item A sharp-edged red shelf extending beyond the photospheric emission in \Ha{} originating from the cool dense shell, although this feature is only starting to become visible on day 50 
    \item A strong, narrow absorption feature at the velocity of the dense shell in \Ha, \Hb, \ion{Na}{1}\,D $\lambda\lambda$5890, 5896, and the \ion{Ca}{2} NIR triplet $\lambda\lambda\lambda$8498, 8542, 8662 increasing in strength with time starting around day 30.
\end{enumerate}

For the first point, we find that the UV luminosity is consistent with models with CSM interaction at all epochs, and not just limited to the early epochs when the narrow emission features are present.
The fact that the shock power in the best-matched model decreases with the lookback time indicates that the ejecta are sweeping through CSM created from decreasing mass-loss rates. 
This is consistent with the light-curve modeling results of \citet{2024Singh} which require extended CSM ($\mathrm{\dot{M}=10^{-4}}$ \masslossunit to $\mathrm{R=10^{16}}$ cm) to match the late-time UV excess.
We detect a boxy \MgII\ emission feature at 66 days, indicating CSM interaction even without the presence of a UV continuum, confirming the second expectation.
As described in the third point, the absorption components in the P~Cygni profiles of the observed spectra are very shallow and slow to develop, especially in \Ha\ which is only clearly identified in the final epoch. This is also notes by \citet{2024Singh}.
The fourth prediction of a sharp red shelf in  \Ha{} is subtle on day 66 and not visible prior to this date. 
At the S/N of our data, we cannot confidently identify it nor rule out its presence. 
Similarly, the narrow high-velocity absorption component, if present, is significantly weaker than predicted in the models.
This could be due to a lower density CSM, asymmetric CSM where the bulk of the CSM is offset from the line of sight, and/or
 the fragmenting of the cool dense shell due to Rayleigh Taylor instabilities which would spread this feature out in the velocity space. 

To explore whether clumping and lowering the velocity of the cool dense shell would provide a better fit to our observations, we calculate a custom CMFGEN model. 
We place the cool dense shell at 8500 \kms. 
At this velocity, the ejecta are optically thick.
This has two effects: first, the material has not been able to cool, so the temperature is higher; and second, it responds much more quickly to the power from the CSM interaction.
However, we found that this change alone was not enough to reproduce our observations and continued to further tune the model.
To better approximate the physical conditions that create the emission from the cool dense shell (i.e., the reprocessing of X-rays from the forward and reverse shock), we add a uniform field of X-rays starting at 8000 \kms\ with a total power of $5\times10^{42}$~ \pwrunit{} instead of inserting power into the cool dense shell directly. 
Given the energy of these X-rays, we add higher energy ions to our model, namely, \ion{C}{3}, \ion{N}{3}, \ion{O}{3}, \ion{Mg}{3}, and \ion{Ne}{3}. 
We also increase the clumping from 1\% to 10\% volume filling factor outside of 8000 \kms, keeping the total mass the same. 

\begin{figure}
    \centering
    \includegraphics[width=\columnwidth]{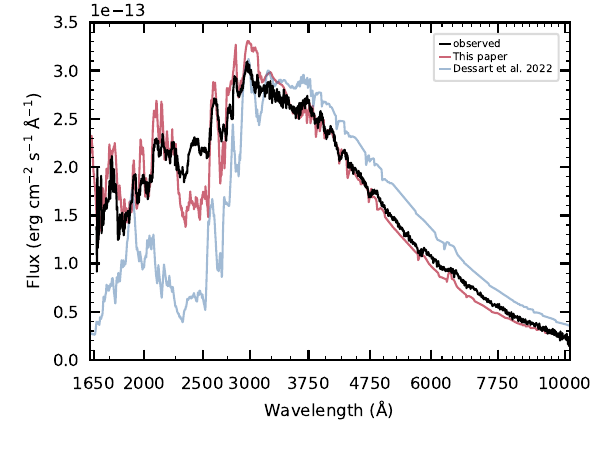}
    \caption{Observed spectrum on day 14 compared to the $L_{\mathrm{sh}}=5\times10^{42}$~ \pwrunit{} model from \citet[][blue]{2022Dessart} and our new model with the cool dense shell at lower velocity, increased clumping, and a more complete atomic model (pink). 
    The new model is able to match the velocity of most of the features and provides a significantly better match to the SED. }
    \label{fig:NewMod}
\end{figure}

In \autoref{fig:NewMod} we compare this model on day 14.6 to our day 14.2 observed spectrum and the best-matching model from \citet{2022Dessart} on day 14.6.
Reducing the velocity of the cool dense shell aligns the absorption and emission.
However, the resulting increase in the temperature of the cool dense shell shifts flux from the optical to the UV, causing the UV to overshoot the observed spectrum and the optical to underestimate the true flux.
The change in power deposition from direct injection into the cool dense shell to an X-ray field reduces the UV flux to align with the observations but does not resolve the low optical flux issue.
The increase in clumping from a volume filling factor of 1\%--10\% increases the optical flux to match the full SED. 
The observed spectrum is very well matched by our custom model, demonstrating the success of this approach. 
However, further improvements could be made to better match the strength of features in the UV, most notably the \ion{Fe}{2} absorption complex in the range 2250--2760~\AA, which is too strong in the models.

    \subsection{CSM Density Profile and Progenitor Mass-loss History}
\begin{figure}
    \centering
    \includegraphics[width=\columnwidth]{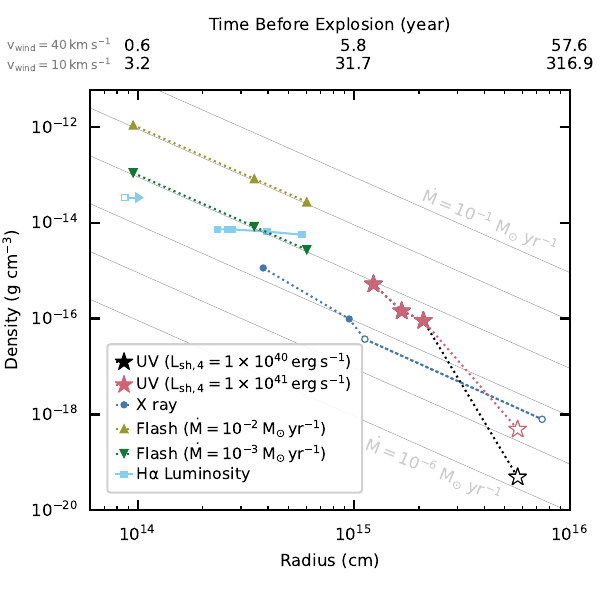}
    \caption{CSM density of SN~2023ixf as a function of the radius assuming a shock velocity of 10,000 \kms.
    The CSM density constraints from this paper are plotted with star symbols.
    The filled symbols are constrained by the overall UV and optical flux. 
    The open stars represent the value from the \MgII\ line, which lies between $L_{\rm sh}=10^{40}$~ \pwrunit\ (black) and $L_{\rm sh}=10^{41}$~ \pwrunit\ (pink).
    Density constraints from flash spectroscopy are shown at the epoch of the first spectrum, day 4, when the majority of the narrow features disappear, and day 7, when all of the narrow features disappear \citep{2023Bostroem2, 2023Jacobson-Galan, 2023Zhang}. Mustard triangles represent a mass-loss rate of $10^{-2}$~ \masslossunit\ and green inverted triangles represent a mass-loss rate of $10^{-3}$~ \masslossunit. 
    Values derived from the \Ha\ luminosity from \citet{2023Zhang} are shown as light-blue solid squares and the value from \citet{2023Bostroem2} is a light-blue open square.
    CSM density from NuSTAR X-ray observations \citep{2023Grefenstette} are plotted as filled blue circles and X-ray observations from {\it Chandra} \citep{2024Chandra} are shown as open blue circles.
    Lines of constant density corresponding to mass-loss rates of $10^{-1}$, $10^{-2}$, $10^{-3}$, $10^{-4}$, $10^{-5}$, and $10^{-6}$ \masslossunit\ are shown as solid gray lines. 
    While there is significant scatter in the density, it is clear that density is decreasing more rapidly than would be expected from constant mass-loss rates down to the density expected for canonical RSG winds.}
    \label{fig:densityHistory}
\end{figure}

Using the best models at each epoch, we derive the density profile of the RSG progenitor of SN~2023ixf, which is given in \autoref{tab:CSMParams} and plotted in \autoref{fig:densityHistory}.
We use the shock luminosity of each model, assuming a shock velocity of $v_{\rm sh}=10,000$ \kms\ to find the CSM density at each epoch from $\rho(r) = L_{\rm sh}/(2 \pi v_{\rm sh}^{3}r^{2})~ \mathrm{g~cm^{-3}}$. 
Note that this is a simplified model that assumes both spherical symmetry and a shock velocity. 
From this, we find that the density decreases more rapidly than a constant wind, going from $\rho=5.2\times10^{-16}$~\densityunit\ at 14 days after explosion to $\rho=4.8\times10^{-20}$~\densityunit\ on day 66.

\begin{figure}
    \centering
    \includegraphics[width=\columnwidth]{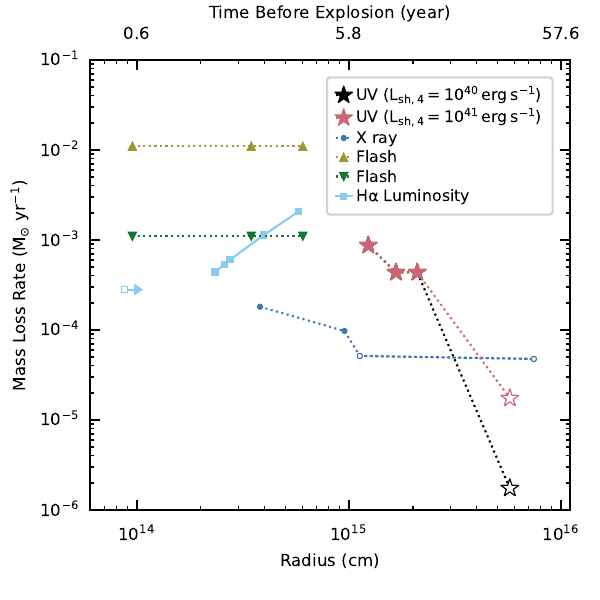}
    \caption{Same as \autoref{fig:densityHistory} for the mass-loss history of SN~2023ixf assuming a shock velocity ($v_{\rm sh}=10,000$ \kms) and the CSM wind velocity ($v_{\rm wind}=55$ \kms). 
    In its last $\sim$30 yr of life, the mass-loss rate increased from a nominal RSG wind of $\sim10^{-6}$ \masslossunit\ to $\sim10^{-2}$ \masslossunit. }
    \label{fig:MassLossHistory}
\end{figure}

If we further use the RSG wind velocity of $v_{\rm wind}=55$ \kms\ measured by \citet{2023Zhang}, we can derive a mass-loss rate from the density and a time of mass loss from the radius (see \autoref{fig:MassLossHistory}).
We find $\dot{M}=8.7\times10^{-4}$~\masslossunit{} 14 days after the explosion, decreasing to $4.4\times10^{-4}$~\masslossunit{} 19 and 24 days after the explosion, and finally arriving at a canonical RSG wind of $1.7\times10^{-6}$~\masslossunit\ at 66 days after the explosion.
This traces the mass loss from elevated rates $\sim7$ yr before the explosion down to nominal RSG rates $\sim35$ yr before explosion.
The clear presence of \MgII\ emission on day 66 indicates that we can detect ejecta-CSM interaction, even with canonical RSG winds of $10^{-6}$~\masslossunit{}.
This mass-loss rate and timing are dependent on the assumed wind velocity. 
If \citet{2023Zhang} measured a wind that had already been radiatively accelerated, the velocity could be lower, which would increase the period of time prior to the explosion probed by these observations and decrease the mass-loss rate. 

The CSM density and mass-loss rates derived in this paper can be combined with other published values to paint a more complete picture of the CSM around the progenitor of SN~2023ixf. 
We plot these densities and mass-loss rates in \autoref{fig:densityHistory} and \autoref{fig:MassLossHistory}.
\citet{2023Grefenstette} use NuSTAR X-ray observations on day 4.4 and day 11 to measure a column density of $N_{\rm H}=26^{+5}_{-7}\times10^{22}$~ $\mathrm{atoms~cm^{-2}}$ and $N_{\rm H}=5.6\pm2.7\times10^{22}$~ $\mathrm{atoms~cm^{-2}}$, respectively.
Additionally, from {\it Chandra} observations on days 13 and 86, \citet{2024Chandra} find a column density of $N_{\rm H}=2.5^{+0.40}_{-0.34}\times10^{22}$~ $\mathrm{atoms~cm^{-2}}$ and $N_{\rm H}=0.36^{+0.22}_{-0.17}\times10^{22}$~ $\mathrm{atoms~cm^{-2}}$, respectively. 
From the column density, the CSM density can be found assuming a density profile from a constant wind and that $N_{\rm H} = \int^{\infty}_{R_{\rm sh}} \rho(r)/m_{p}\, dr$. With these assumptions, the density at each epoch is $\rho=(N_{\rm H}m_{P})/R_{\rm sh}$.
From the best-matching models of the observed flash features between 1 and 7 days, \citet{2023Bostroem2}, \citet{2023Jacobson-Galan}, and  \citealt{2023Zhang} infer a mass-loss rate of $10^{-3}-10^{-2}$~ \masslossunit\ from the CMFGEN models of \citet{2017Dessart} and \citet{2023Jacobson-Galan}.
As the line emission in these models is dependent on the density and mass loss is a parameterization of this, we use the model wind velocity of $v_{\rm wind}=50$ \kms\ to find the density rather than the measured wind velocity.  
We plot the density derived from flash features at the time of the first spectrum (day 1.1), day 4 when many of the narrow features disappear, and day 7 when the \Ha\ narrow component disappears.
\citet{2023Bostroem2} and \citet{2023Zhang} use the \Ha\ luminosity to calculate the CSM radius and density. 
\citet{2023Bostroem2} find $R_{\rm CSM} \gtrsim 8.7\times10^{13}$ cm and $\rho_{\rm CSM}=3.4\times10^{-14}$~ \densityunit.
\citet{2023Zhang} determine $R_{\rm CSM}=2.33$, 2.59, and $2.74\times10^{14}$ cm and $\rho_{\rm CSM}=7.41$, 7.26, and $7.31\times10^{-15}$~ \densityunit.

These combined measurements show that the mass-loss rate of the progenitor of SN~2023ixf was relatively low until quite close to the time of core collapse ($R\lesssim5\times10^{15}$ cm or $t\lesssim33$ yr prior to explosion), at which point it increased to the large densities inferred from flash-feature observations. 
However, this figure also highlights that the systematics, uncertainties, and simplifying assumptions in these different techniques lead to 1 order of magnitude uncertainties in mass-loss rates and densities.

\section{Summary}\label{sec:summary}
We present UV observations of SN~2023ixf 14--66 days after the explosion.
SN~2023ixf is UV bright, with continuum emission through day 24.  
Its UV spectrum  shows weaker metal absorption than other Type II supernovae that do not exhibit definitive signs of CSM interaction. 
In the optical, the \Ha\ P~Cygni profile is slow to develop and has a shallow absorption component, which we infer is due to the additional emission from CSM interaction.  
On day 66, the \MgII\ profile and \Ha\ emission component have a similar extent in velocity space, but extremely different profile shapes, resulting from the lack of UV continuum from the supernova ejecta.
We examine the contribution of different species to the UV spectrum of SN~2023ixf on days 14 and 19, finding that the majority of the absorption in the UV spectrum is due to iron, nickel, and magnesium at both epochs, with additional contributions by chromium at the later epoch.

We compare SN~2023ixf to CMFGEN models of supernova ejecta with power from CSM interaction \citep{2022Dessart}. 
Impressively, even though these models were not made for a supernova like SN~2023ixf, which interacted with dense CSM during its early evolution, they match the overall evolution and the presence of individual features.
It is likely that the cool dense shell in SN~2023ixf is broader and centered at a lower velocity (8500 vs. 11,700 \kms) and higher temperature than the models.
This is confirmed by a custom model that places the cool dense shell at 8500 \kms.
Additionally, the predicted presence of the \MgII\ emission after 50 days is confirmed, demonstrating that SN~2023ixf is interacting with CSM at all epochs through day 66.

We visually identify the model that best matches our observed spectrum at each epoch, finding that the power from CSM interaction decreases with time.
From this, we conclude that at larger radii, there is lower density CSM, suggesting that the mass-loss rate increased rapidly just prior to the explosion.  
While the flash features favor a confined CSM, these disappear after a week, limiting the radius out to which the mass-loss history can be traced by this method. 
Our UV observations are sensitive to much lower CSM densities, and we use them to trace the CSM density to significantly larger radii and the mass-loss history to earlier epochs.
With a shock velocity of 10,000 \kms,  our observations probe the CSM density between a radius of $1.2\times10^{15}$ cm to $5.2\times10^{15}$ cm and show the density evolving from $5.2\times10^{-16}$~ \densityunit\ to $4.8\times10^{-20}$~ \densityunit\ over this range.
Using the wind velocity of 55 \kms, we find that the mass-loss rate decreases from $8.7\times10^{-4}$~ \masslossunit\ in our first observation to $1.7 \times 10^{-6}$~ \masslossunit\ in our last observation, showing the dramatic change in mass-loss rates 33--7~yr before the explosion.
Additionally, the mass-loss rate derived from the day 66 spectrum shows that 33~yr before the explosion, the RSG progenitor's mass-loss rate was consistent with a quiescent RSG wind \citep{2020Beasor}.

We also compare SN~2023ixf with other supernovae having UV spectra around days 10 and 20.
These objects cover a range of UV fluxes, similar to the range spanned by the CMFGEN models with varying levels of CSM interaction. 
This suggests that we can use the UV to sensitively distinguish varying levels of CSM interaction. 
Around day 10, we see that \MgII\ is clearly present in absorption in the spectra with the most CSM interaction. 
Often appearing next to it is \ion{Fe}{2}; this can be seen as a distinct feature in SN~2023ixf and SN~2022wsp, and it may be present in the other supernovae either blended with \MgII\ owing to its formation at higher velocities or as a weaker feature.
\MgII\ and \ion{Fe}{2} are weaker around day 20 in all supernovae. 
Additionally, the apparent emission line at 1924\,\AA\ is caused by a window of lower iron absorption, rather than emission from an individual species, demonstrating the complexity of associating a single species with an individual feature in UV spectra.

Based on the shape and width of the \MgII\ emission in the day 66 spectrum, we conclude that this emission is from the cool dense shell, which is well described by a thick shell with a Gaussian density profile: FWHM $\approx 7700$ \kms and an inner radius velocity of $\sim -6500$ \kms.
We model the emission with an asymmetric Gaussian shell, finding that the profile is consistent with more of the observed emission coming from the blue side than the red side.
Additionally, in other supernovae, this feature is either symmetric or asymmetric with a higher blue side, leading us to conclude that the profile shape is due to attenuation of the far side of the shell, likely from opacity in the shell and ejecta at early times and dust attenuation in the later-time spectra (e.g. SN~1998S at 240d). 

Finally, we compile the density and mass-loss measurements of SN~2023ixf from the literature using a consistent shock velocity of 10,000 \kms\ and a wind velocity of 55 \kms. 
We show that the density profile from $9.5\times10^{13}$ cm to $7.5\times10^{15}$ cm (0.5--42~yr before explosion) decreases from $1.1\times10^{-12}$ \densityunit\ to $4.9\times10^{-20}$ \densityunit.
With a wind velocity of 55 \kms, this corresponds to a change in mass-loss rate of $\sim10^{-2}$ \masslossunit\ to $\sim10^{-6}$ \masslossunit. 
This shows that over the final $\sim 40$~yr of the progenitor's life, the mass-loss rate went from that of a quiescent RSG to an extremely high mass-loss rate.
The densities and mass-loss rates of these different techniques span about an order of magnitude, highlighting the uncertainties and systematic errors in our methods, even when some consistent assumptions are made.

SN~2023ixf is the best-observed Type II supernova since SN~1987A, providing us with a Rosetta Stone to interpret other observations and against which to benchmark theoretical predictions.
Although relatively few UV observations of Type II supernovae exist (and even fewer time series), it is clear that this wavelength range is rich in information about the temperature, composition, and dynamics of the CSM and the ejecta. 
It also probes the pre-supernova mass-loss history, tracing it back to quiescent RSG winds 40 years before explosion and can provide insights into the role of CSM interaction in the diversity of optical properties that we observe in Type II supernovae. 
In particular, time series which look for the presence of \MgII\ emission for all Type II supernovae and trace its evolution to even later epochs are warranted inform our understanding of mass loss in RSGs, especially in the years leading up to explosion. 
The observations of SN~2023ixf provide a critical link between early- and late-time UV observations, and also between `normal' Type II supernovae and strongly-interacting Type II supernovae.

\movetabledown=5cm
\begin{rotatetable}
\begin{deluxetable*}{lcccccc}
\tablecaption{Properties of Supernovae with UV Spectra Analyzed in This Paper \label{tab:compSN}}
\tablehead{\colhead{Name} & \colhead{Spectra Epochs} &\colhead{Distance Modulus} & \colhead{Explosion Epoch} & \colhead{MW $E(B-V)$} & \colhead{Host $E(B-V)$} & \colhead{References}\\ 
                      & \colhead{(day)}             & \colhead{(mag)}   & \colhead{(JD)}            &     \colhead{(mag)} &\colhead{(mag)}& }

\startdata
     SN 1979C    &  16, 24, 3, 34, 41, 48, 59, 69, 77, 82, 120    & 31.04 $\pm$0.17      & 2443970.5  $\pm$ 8       & 0.0226               & 0.16 $\pm$ 0.05   & (A, B--H)\\
     SN 1980K    &  10, 11, 13, 20, 36, 45, 73                    & 29.44 $\pm$ 0.21 & 2444537.8$^{+}$  $\pm$ 10           & 0.2930               & 0.00              & (A, I, J, K)\\
     SN 1987A    &  8, 9, 11, 21, 32, 42, 51, 62, 68, 102, 201    & 18.55 $\pm$ 0.05 & 2446849.62 $\pm$ 1            & 0                    & 0.19 $\pm$ 0.02   & (L--T)\\
     SN 1993J    &  18, 19, 174, 670                              & 27.80 $\pm$ 0.10 & 2449074.0  $\pm$ 0.8          & 0.0670               & 0.10 $\pm$ 0.11   & (A, U, V, W, X, Y)\\
     SN 1998S    &  17, 28, 76, 241, 490                          & 31.18 $\pm$ 0.38 & 2450871.7  $\pm$ 3.5          & 0.0202 $\pm$ 0.0009  & 0.20 $\pm$ 0.02   & (A, Z, AA, BB, CC, DD)\\ 
     SN 1999em   &  13                                            & 30.34 $\pm$ 0.07 & 2451475.9$^{\dagger}$ $\pm$ 2 & 0.03486 $\pm$ 0.0003 & 0.10 $\pm$ 0.05   & (A, EE, FF, GG, HH)\\
     SN 2005cs   &  11                                            & 29.39 $\pm$ 0.47 & 2453549.5$^{*}$  $\pm$ 1          & 0.0307 $\pm$ 0.0018  & 0.015             & (A, GG, LL, MM, NN, OO) \\
     SN 2021yja  &  10, 21                                        & 31.85 $\pm$ 0.45 & 2459464.9  $\pm$ 0.06         & 0.019 $\pm$ 0.085    & 0.085 $\pm$ 0.015 & (A, SS, TT)\\
     SN 2022wsp  &  10, 20                                        & 31.99 $\pm$ 0.16 & 2459855.08 $\pm$ 0.0484       & 0.05                 & 0.30              & (A, UU)\\
     SN 2022acko &  7, 19                                         & 31.39 $\pm$ 0.33 & 2459918.7  $\pm$ 1            & 0.0259 $\pm$ 0.0002  & 0.03 $\pm$ 0.01   & (A, VV, WW)\\
\enddata
\tablecomments{$^{+}$ SN~1980K does not have a reliable explosion epoch. We adopt an explosion epoch 7 days prior to an observed spectrum on 1980 November 1 which shows broad H$\alpha$ emission. $^{\dagger}$See also \citealt{2006Dessart}, $^{*}$See also \citealt{2008Dessart}. 
(A): \citealt{2011Schlafly}. 
(B): \citealt{2015Gall}. 
(C): \citealt{1996Ferrarese}. %
(D): \citealt{1982Barbon}. 
(E): \citealt{1980Panagia}.
(F): \citealt{1982Palumbo}.
(G): \citealt{1982Benvenuti}.
(H): \citealt{1999Fesen}.
(I): \citealt{2019VanDyk}. 
(J): \citealt{1982Barbon2}. 
(K): \citealt{1982Pettini}.
(L): \citealt{2000Romaniello}. 
(M): \citealt{1987Castagnoli}. 
(N): \citealt{1996Scuderi}. 
(O): \citealt{1987Wamsteker}.
(P): \citealt{1987Cassatella}.
(Q): \citealt{1987Cassatella2}.
(R): \citealt{1987Kirshner}.
(S): \citealt{1988Panagia}.
(T): \citealt{1995Pun}.
(U): \citealt{1994Freedman}. 
(V): \citealt{1994Lewis}. 
(W): \citealt{2014Ergon}. 
(X): \citealt{1994Jeffery}.
(Y): \citealt{2005Fransson}.
(Z): \citealt{1997Willick}. 
(AA): \citealt{2023Bostroem2}. 
(BB): \citealt{2000Leonard}. 
(CC): \citealt{2001Lentz}.
(DD):\citealt{2005Fransson}.
(EE): \citealt{2003Leonard}. 
(FF): \citealt{2003Elmhamdi}. 
(GG): \citealt{2017Silverman}. 
(HH): \cite{2000Baron}.
(II): \citealt{1988Tully}. 
(JJ): \citealt{2006Tsvetkov}. 
(KK): \citealt{2008Gal-Yam}.
(LL): \citealt{2018Sabbi}. 
(MM): \citealt{2006Pastorello}. 
(NN): \citealt{2007Baron}. 
(OO): \citealt{2007Brown}.
(PP): \citealt{2012Zhang}. 
(QQ): \citealt{2011Stoll}. 
(RR): \citealt{2014Fransson}. 
(SS): \citealt{2022Hosseinzadeh}. 
(TT): \citealt{2022Vasylyev}.
(UU): \citealt{2023Vasylyev}. 
(VV): \citealt{2021Anand}. 
(WW): \citealt{2023Bostroem1}. 
}
\end{deluxetable*}
\end{rotatetable}
\clearpage

\begin{acknowledgments}

We thank the referee for their thoughtful comments.
This research is based on observations made with the NASA/ESA {\it Hubble Space Telescope} obtained from the Space Telescope Science Institute, which is operated by the Association of Universities for Research in Astronomy, Inc., under NASA contract NAS 5-26555. These observations are associated with programs GO-17313 and GO-17205.
Huge thanks to our {\it HST} program coordinator Alison Vick and our contact scientist Dan Welty for assistance in planning and executing the observations.
The data described here may be obtained from the MAST archive at
\dataset[doi:10.17909/787k-c897]{https://dx.doi.org/10.17909/787k-c897}.

K.A.B. is supported by an LSST-DA Catalyst Fellowship; this publication was thus made possible through the support of grant 62192 from the John Templeton Foundation to LSST-DA.
A.V.F. is grateful for financial assistance from the Christopher R. Redlich Fund and many other donors. Time domain research by D.J.S. and team is supported by NSF grants AST-1908972 and 2108032, and by the Heising-Simons Foundation under grant No. 20201864. 
Research by Y.D., S.V., N.M.R, E.H., and D.M. is supported by NSF grant AST-2008108.

\end{acknowledgments}
\vspace{5mm}
\facilities{HST(STIS)}


\software{astropy \citepalias{astropy_collaboration_astropy_2013, astropy_collaboration_astropy_2018, astropy_collaboration_astropy_2022}, CMFGEN \citep{1998Hillier, 2013Dessart, 2019Hillier}, MatPLOTLIB \citep{hunter_matplotlib_2007},  NumPy \citep{harris_array_2020}, Scipy \citep{virtanen_scipy_2020}, stistools \citep{2019Sohn}}

\bibliography{sn2023ixf_UV}{}
\bibliographystyle{aasjournal}
\end{document}